# Harvesting in a resource dependent age structured Leslie type population model

Rui Dilão[1], Tiago Domingos[1] and Elman M. Shahverdiev[1,2]

*1) Grupo de Dinâmica Não-Linear, Instituto Superior Técnico, Av. Rovisco Pais, 1049-001 Lisboa, Portugal.*

*2) Institute of Physics, 33, H. Javid Avenue, 370143 Baku, Azerbaijan.*

**Short Title:** Harvesting in a resource dependent model.

**Corresponding author:**

Rui Dilão

Grupo de Dinâmica Não-Linear, Instituto Superior Técnico

Av. Rovisco Pais, 1049-001 Lisbon, Portugal

rui@sd.ist.utl.pt

Phone (351)218417617; Fax (351)218419123.

## Abstract

We analyse the effect of harvesting in a resource dependent age structured population model, deriving the conditions for the existence of a stable steady state as a function of fertility coefficients, harvesting mortality and carrying capacity of the resources. Under the effect of proportional harvest, we give a sufficient condition for a population to extinguish, and we show that the magnitude of proportional harvest depends on the resources available to the population. We show that the harvesting yield can be periodic, quasi-periodic or chaotic, depending on the dynamics of the harvested population. For populations with large fertility numbers, small harvesting mortality leads to abrupt extinction, but larger harvesting mortality leads to controlled population numbers by avoiding over consumption of resources. Harvesting can be a strategy in order to stabilise periodic or quasi-periodic oscillations in the number of individuals of a population.

# 1. Introduction

In natural environments and in the laboratory the number of individuals of a population change in time in different ways. For example, in fishes, Dixon, Milicich & Sugihara [1] reported the erratic replenishment of open marine ecosystems, and, in laboratory experiments with insects, Costantino et al. [2] observed quasi-periodic oscillations and transitions to chaos by setting high the adult mortality of *Tribolium*. These observations show that the number of individuals of a population can have large amplitude oscillations both in nature and laboratory. Models presenting qualitatively this type of behaviour are density dependent unstructured population models [3], density dependent age structured population models [2], [4]-[6] and, more naturally, resource dependent age structured population models [7]. How harvesting affects the temporal dynamics of populations in the context of these models is an open question.

For a logistic growth population model, [3], the effect of harvesting has been analysed by Beddington & May [8] and, from the point of view of bio-economics, by Clark [9]. In logistic approaches without harvesting, asymptotic population numbers have always a stable steady state, [9] and [10], and the effect of resources is modelled by a carrying capacity parameter associated with a mass conservation law, [11]. Getz & Haight [12] have analysed the harvesting of density dependent age structured population models in the case where the attractor of the dynamics is a period-1 fixed point. Harvesting with constant depletion in one-dimensional maps with one critical point has been analysed by Schreiber [13].

In density dependent models, resources are not explicitly modelled. They are considered implicitly, by assuming that they are low when population numbers are high and vice-versa. Survival and/or fertility are then taken to decrease with increasing population numbers, [5] and [6]. This implicitly assumes that resources have a very fast dynamics, and instantaneously adjust to the population level. We do not expect this to be the most frequent case in nature. Although resources will decrease with increasing population densities, this is not an instantaneous process, and there will be situations of high population densities with high resources and low population densities with low resources [14]. To encompass all these possibilities, it becomes necessary to model resources explicitly.

On the other hand, harvesting effects can be seen as a possible approach to analyse the effect of predation in species that are naturally incorporated in trophic chains: harvesting corresponds to a specific type of predation and resources correspond to the supporting species.

Here, we analyse the time evolution of a Leslie type resource dependent age structured population exposed to harvesting. Populations are considered age structured and the resources are explicitly incorporated into the model. The main question addressed in this paper is the following: If a population has periodic, almost-periodic or chaotic time behaviour in the number of individuals, what is the effect of harvesting in the dynamics of the population?

In section 2, we analyse a model for a resource dependent population and we describe some of its dynamic features. In section 3, we introduce harvesting with constant mortality, and we compare the dynamics of the models with and without harvesting. One important conclusion that derives from this comparison is that, for low harvesting mortality, the model with harvesting is equivalent to the model without harvesting for a different choice of parameter values. When the harvesting mortality is high, then the new model becomes structurally different from the initial model. In section 4, we make the numerical analysis of the harvesting model map for several choices of bifurcation parameters. In particular, we analyse the harvesting yield and the conditions leading to the survival and extinction of a population. Finally, in section 5, we summarize the main conclusions of the paper.

## 2. The basic model for a resource dependent population

To analyse the effect of harvesting on a population, we first consider a basic model describing the time evolution of a resource dependent age structured population. In principle, the effect of harvesting depends on the intrinsic features of the dynamics of the non harvested population.

Let $N_i^n$ represent the number of individuals of a population with age $i$ at time $n$, and let $R^n$ represent a generic resource at discrete time $n$. A model that describes the time evolution of the population depending on a generic resource is, [7],

$$\begin{cases} N_1^{n+1} = \sum_{i=2}^{k} e_i N_i^n \\ N_i^{n+1} = \alpha_{i-1}(R^n) N_{i-1}^n, \ i=2,...,k \\ R^{n+1} = f(R^n)\phi(N^n) \end{cases} \quad (2.1)$$

where $e_i$ is the fertility coefficient of age class $i$, $N_1^n$ represents the new-born age class, and $N^n = \sum_{i=1}^{k} N_i^n$ is the total number of individuals in the population. The index $k$ represents the oldest reproductive age class, and the phase space of system (2.1) has dimension $k+1$. In the following, we assume that $k$ is fixed and $e_k > 0$. The coefficients $\alpha_{i-1}(R^n)$ are resource dependent transition probabilities between age classes $i-1$ and $i$, and relate to death rates $d_{i-1}(R^n)$ through $\alpha_{i-1}(R^n) = 1 - d_{i-1}(R^n)$. As an approximation, we consider that adult fertilities are resource independent. The function $\phi(N^n)$ is continuous and decreasing, with $0 \leq \phi(N^n) \leq 1$. The last equation in (2.1) assumes that resources are reduced proportionally to a function of the number of encounters with consumers. To study the survival of a population it is enough to incorporate into the model only reproductive age classes, [15] and [6].

In the absence of consumers, the resource dynamics in (2.1) is,

$$R^{n+1} = f(R^n)\phi(0) = f(R^n) \quad (2.2)$$

where $\phi(0) = 1$. If the resource dynamics (2.2) has a stable period-1 fixed point $R^n = K$, with $f(R^n)$ a monotonic increasing function of $R^n$, the point $(R = K, 0, ..., 0)$ is always a fixed point of map (2.1). The constant $K$ is the carrying capacity.

As in the case of resource independent Leslie maps, [6], we define the resource dependent inherent net reproductive number of the population $G(K)$ by,

$$G(K) = e_2\alpha_1(K) + ... + e_k\alpha_1(K)...\alpha_{k-1}(K) \qquad (2.3)$$

It can be shown that, [7], if $G(K) < 1$ , then the period-1 fixed point $(R = K, 0, ..., 0)$ of map (2.1) is stable. If $G(K) > 1$, then (2.1) has a second period-1 equilibrium point $(R^*, N_1^*, ..., N_k^*)$, with $0 < R^* < K$ and $N_i^* > 0$, for $i = 1, ..., k$ . Under the hypothesis that the functions $\alpha_i(R), \phi(N), f(N)$ are continuous, differentiable and monotonic, the map (2.1) is invertible (diffeomorphism) in the set $(R > 0, N_1 \geq 0, ..., N_k \geq 0)$ . For parameter values such that $G(K) > 1$ but $G(K)$ is close to 1, then the fixed point $(R^*, N_1^*, ..., N_k^*)$ is stable, and the map (2.1) is structurally stable in the positive quadrant of phase space. The structural stability property of map (2.1) implies that the topological properties of the solutions of (2.1) in phase space do not change under small perturbations of the chosen specific functional forms for $f$ , $\alpha_i$ and $\phi$ . For $G(K) = 1$, the map (2.1) is not structurally stable and has a transcritical or fold bifurcation, [7].

Numerical analysis for a choice of the functions in (2.1) shows that, for some value of the parameter $G(K)$, with $G(K) \gg 1$, the steady state $(R^*, N_1^*, ..., N_k^*)$ loses stability by flip (period doubling) or discrete Hopf bifurcations, and quasi-periodic behaviour in time appears. For increasing values of $G(K)$ and in the case where $(R^*, N_1^*, ..., N_k^*)$ loses stability by a period doubling bifurcation, the cascade of bifurcations is interrupted by a Hopf bifurcation. The transition from periodic to non periodic time behaviour always occurs by a discrete Hopf bifurcation from a stable period $q \geq 1$ point.

After the Hopf bifurcation, the attractor of map (2.1) is an invariant circle in phase space. For the detailed derivation of the properties of map (2.1) see [7].

In figure 1, we depict in grey the bifurcation diagram of map (2.1), for the choice of functions:

**Error! Reference source not found.**
$$\begin{cases} f(R) = \dfrac{K\beta R}{R(\beta - 1) + K} \\ \alpha_i(R) = \dfrac{\gamma_i R}{\gamma_i R + 1} \\ \phi(N) = e^{-\mu N} \end{cases}$$

(2.4)

where $\beta$ is the discrete time intrinsic growth rate of the resources (in the absence of consumers, resources are assumed to evolve in time with a discrete logistic growth), $\mu$ is a scaling constant, and $\gamma_i > 0$ are constants associated to a birth-and-death process with probability $\alpha_i(R) = \gamma_i R / (\gamma_i R + 1)$, [16]. The resource dynamics evolves according to a Beverton-Holt dynamics, and resources are reduced proportionally to the number of encounters with consumer, thus having a Poisson or Ricker form ([6, p. 12-13]).

Map (2.1) with the functions (2.4) is our reference model to analyse the effect of harvesting in resource dependent populations with periodic and non-periodic time behaviour.

## 3. Harvesting with constant mortality

The simplest way to consider the effect of harvesting in model map (2.1) is to deplete individuals from each age class. As in the harvesting process it is difficult to control age class selection, we suppose that, in each time interval [n,n+1], the total depletion of the population is $\tilde{N}^n$, where $0 \leq \tilde{N}^n \leq N^n$. We define the harvesting mortality of the population as $\tilde{N}^n / N^n = \delta \leq 1$, and the harvesting yield as $\tilde{N}^n = \delta N^n$. For $\delta = 0$ there is no harvesting.

Assuming that all the age classes are well mixed in the population, the depletion of each age class is $\tilde{N}^n(N_i^n / N^n) = \delta N_i^n$, where $(N_i^n / N^n)$ is the fraction of individuals of the population with age $i$. As only non harvested individuals will affect the resources, introducing harvesting into (2.1), we obtain the new map,

$$\begin{cases} N_1^{n+1} = \sum_{i=2}^{k} e_i N_i^n \\ N_i^{n+1} = \tilde{\alpha}_{i-1}(R^n) N_{i-1}^n, \ i = 2,...,k \\ R^{n+1} = f(R^n)\phi(N^n - \tilde{N}^n) = f(R^n)\phi(N^n(1-\delta)) \end{cases} \qquad (3.1a)$$

where,

$$\tilde{\alpha}_{i-1}(R^n) = \begin{cases} \alpha_{i-1}(R^n) - \delta, \ if \ \alpha_{i-1}(R^n) - \delta > 0 \\ 0, \ if \ \alpha_{i-1}(R^n) - \delta \leq 0 \end{cases} \qquad (3.1b)$$

Map (3.1) describes the effect of harvesting in a resource dependent population. For $\delta = 0$, map (3.1) reduces to map (2.1), and the problem of harvesting with constant mortality can be analysed through the comparison between the dynamics of maps (2.1) and (3.1). Clearly, the map (3.1) should be understood as a perturbation of map (2.1). Note, however, that the map (3.1) is no more invertible in the set $(R > 0, N_1 \geq 0, ..., N_k \geq 0)$. By (3.1b), it is invertible in the set $(R > \max_i \alpha_i^{-1}(\delta), N_1 \geq 0, ..., N_k \geq 0)$.

Now, we derive the condition for the existence of a non-zero fixed point of equation (3.1). Assuming that $R^n > \max_i \alpha_i^{-1}(\delta)$, for every $n \geq 0$, introducing the change of co-ordinates $M_i^n = (1-\delta)N_i^n$, map (3.1) reduces to the map (2.1), with $\tilde{\alpha}_{i-1}(R^n)$ monotonic and increasing functions of $R^n$. As in the case of map (2.1), the condition for the non-extinction of the population is,

$$G(K, \delta) = e_2(\alpha_1(K) - \delta) + ... + e_k(\alpha_1(K) - \delta)...(\alpha_{k-1}(K) - \delta) > 1 \qquad (3.2)$$

provide that $\delta < \min_i \alpha_i(K)$, and the resource coordinate of the stable stationary state, say $R = \overline{R}$, obeys to $\overline{R}^n > \max_i \alpha_i^{-1}(\delta)$. Therefore, for a sufficiently small harvesting mortality, the map (2.1) has a non zero stable steady state in phase space, and harvesting simply decreases the inherent net reproductive rate of a population.

For larger harvesting mortalities, by (3.1b), if, for some fixed $n \geq 0$, $R^n \leq \min_i \alpha_i^{-1}(\delta)$, at time $n+1$, $N_i^{n+1} = 0$, for $i \geq 2$, and the reproductive age classes extinguish. If, for some fixed $n \geq 0$, $\min_i \alpha_i^{-1}(\delta) < R^n \leq \max_i \alpha_i^{-1}(\delta)$, some of the age classes will die out at time $n+1$. Therefore, a sufficient condition for the extinction of a population by harvesting is that, for some $n \geq 0$, $\delta \geq \min_i \alpha_i(R^n)$.

If the map (3.1) is away from the bifurcation value $G(K,\delta)=1$, and has only one or two fixed points, the orbits in phase space of map (3.1), with $\delta$ sufficiently small, are qualitatively similar to the orbits of map (2.1). This is a consequence of the structural stability property of map (2.1) and by the nature of the functional perturbation introduced by the harvesting effect. The smallness of the harvesting mortality necessary to maintain the same qualitative dynamic features of both maps depends on the parameter values and consequently on dynamics of the non harvested population.

If the population has an equilibrium steady state $N^*$, then the harvesting yield is $\tilde{N}^n = \delta N^*$ ([8], [9]). If the map (3.1) has a periodic, quasi-periodic or a chaotic dynamics, the harvesting yield $\delta N_i^n$ can be periodic, quasi-periodic or chaotic, provided the phase space stable attractors obey to the condition $R^n > \max_i \alpha_i^{-1}(\delta)$, for every $n > 0$.

If $\delta \geq \alpha_1(K)$, $(K \leq \alpha_1^{-1}(\delta))$, then, after $k-1$ iterations, the population becomes extinct, the map (3.1) is no longer a diffeomorphism and the fixed point $(R = K, 0, ..., 0)$ is stable. If, for some $j > 1$, $\delta \geq \alpha_j(K)$, but $G(K,\delta) > 1$, harvesting extinguishes all the age classes $i \geq j+1$. Consequently, the disappearance of older reproductive age classes is an indication of over exploitation of a population. So, to avoid over exploitation, and for the population to maintain all its age classes with non-zero values, we must have $\delta < \min_i \alpha_i(K)$.

The bifurcation structure of map (3.1) is complex and depends on a large number of parameters. However, we can fix all the parameters except one, and analyse the variations in the dynamics of map (3.1) when we change one of the parameter. For example, solving the inequality (3.2) for $e_2$, and if $\delta < \min_i \alpha_i(K)$, we obtain,

$$e_2 > \frac{1}{\alpha_1(K) - \delta} - e_3(\alpha_2(K) - \delta) - ... - e_k(\alpha_2(K) - \delta)...(\alpha_{k-1}(K) - \delta) \qquad (3.3)$$

Hence, if the mortality coefficient $\delta$ increases, in order to avoid extinction, the fertility parameter $e_2$ should also increase. This holds whenever each $e_i$ is taken as a bifurcation parameter for the map (3.1).

## 4. Qualitative analysis of the harvesting map

In figure 1, we compare the bifurcation diagrams of maps (2.1) and (3.1) describing a resource dependent population with (black diagram) and without (grey diagram) harvesting. We have taken three age classes ($k = 3$) for the model functions (2.4), and we have chosen $e_2$ as a bifurcation parameter.

For the parameter values of figure 1, the non harvested population has a non-zero steady state if $G(K) = e_2 \alpha_1(K) + e_3 \alpha_1(K) \alpha_2(K) > 1$. As $e_2$ is the bifurcation parameter, the non-zero steady state exists if $e_2 > 0.34$. By (3.2) and harvesting mortality $\delta = 0.2$, the harvested population only survives if the fertility coefficient $e_2$ is such that $e_2 > 0.842$. However, for larger fertility coefficient $e_2 > 3.46$ and the same constant harvesting mortality, the population becomes extinct. In this case, extinction does not occur necessarily from setting high the mortality parameter. From the technical point of view, extinction is due to the fact that, for some $n$, we have $R^n \leq \max_i \alpha_i^{-1}(\delta)$ (see section 3), which means that the increase of the harvesting mortality induces a change in the phase space attractors and the condition $R^n > \max_i \alpha_i^{-1}(\delta)$ is violated.

Comparing the grey and the black bifurcation diagrams in figure 1, we conclude that, for small harvesting mortality and fertility coefficients, the dynamics of the harvested population is qualitatively similar to the dynamics of the non harvested population with lower fertility coefficients. Decreasing $\delta$, both bifurcation diagrams become more and more similar. On the other hand, in populations with one steady state, condition (3.2) establishes that the effect of harvesting with constant mortality is equivalent to the decrease of the probability of transition between age classes. This numerical experiment corroborates the exact results of section 3.

Also, from figure 1, and for large fertility coefficient $e_2 \in [2.5, 3]$, harvesting can be a strategy or a control process in order to stabilise periodic or quasi-periodic oscillations in the number of individuals of a population. This case can be particularly important in population communities that belong to a trophic chain. For example, we can not exclude the effect of inducing erratic time behaviour by decreasing predation or harvesting.

Another important parameter in the analysis of maps (2.1) and (3.1) is the carrying capacity of the resources. In figure 2, we show the bifurcation diagrams of both maps when we vary the carrying capacity parameter $K$.

Comparing the bifurcation behaviour of maps (2.1) and (3.1) as a function of the carrying capacity parameter $K$, the obvious conclusions is that harvesting can have the effect of a control strategy to avoid erratic time behaviour in populations, figure 2. However, for small carrying capacity values, harvesting can lead to abrupt extinction. As already discussed, this bifurcation associated to abrupt extinction occurs when the fixed points of map (3.1) obeys to the condition $R^n = \max_i \alpha_i^{-1}(\delta)$, for every $n \geq 0$.

We now analyse the behaviour of the harvesting yield as a function of mortality parameter $\delta$. We take map (3.1) for the parameters of figure 1 with fixed values of $e_2$ and we vary the mortality coefficient $\delta$. For parameter values where the map (3.1) with $\delta = 0$ has periodic or quasi-periodic behaviour in phase space, the harvesting yield can be periodic or quasi-periodic in time, figure 3a)-3c). Extinction occurs due to two mechanisms. In one case, extinction occurs when inequality (3.2) is violated (figures 3a)-3c)), but it can also occurs when, for some $n$, $R^n \leq \max_i \alpha_i^{-1}(\delta)$, (figure 3c)). From the numerical simulation in figure 3c), it follows that for populations with intrinsic chaotic behaviour, small harvesting mortality can lead to extinction.

In the harvesting process, the number of individuals that survive decreases as the harvesting mortality increases. The maximum threshold values of the harvesting mortality leading to extinction, $\delta_{extin}$, are calculated from (3.2). For the parameter values of figure 3, we have: a) $e_2 = 1.8$, $\delta_{extin} = 0.448$; b) $e_2 = 3.2$, $\delta_{extin} = 0.618$; and c) $e_2 = 3.46$, $\delta_{extin} = 0.637$, in agreement with the bifurcation diagrams of figure 3a)-3c). Maximum temporal mean harvesting yield seems to occur only outside the erratic regions, when the harvesting yield is periodic, figure 3a), 3d) and 3e).

In figure 3c), where $e_2 = 3.46$, $G(K, \delta)$ and the fertility number $\sum_{i=2}^{k} e_i$ are large, but for very small harvesting mortality the population becomes extinct. For larger values of $\delta$, the population survives and the maximum admissible harvesting is given by (3.2). From the dynamical systems point of view this sudden extinction for small harvesting mortality, and survival for higher mortality, depends on the characteristics of the orbits of map (3.1) in phase space. In these situations, the map (3.1) has a random attractor in phase space ([7]) and the time evolution of the number of individuals in the population is erratic. From the biological point of view, if the behaviour of resources is erratic in time and, for some $n$, available resources reaches such low values ($R^n \leq \max_i \alpha_i^{-1}(\delta)$), that the only way to prevent over consumption of resources, ensuring survival, is to increase the harvesting mortality: Increasing harvesting mortality gives more resources to the surviving reproductive individuals of the population.

In figure 4, we show the regions of survival and extinction as a function of the harvesting mortality $\delta$ and fertility coefficient $e_2$ for map (3.1) with $k = 3$. The upper curve is given by the threshold of extinction defined by $G(K, \delta) = 1$ and, in the limit $e_2 \to \infty$, is asymptotic to the line defined by $\delta = \alpha_1(K)$. The lower curve is calculated numerically and depends on the structure of the attractors of map (3.1). In the limit $e_2 \to \infty$, the lower curve is asymptotic to the lines $\delta = \alpha_1(K)$ and $\delta = 0$.

Several numerical experiments were performed with map (3.1) for different parameter values of those of figures 1, 2, 3 and 4. In all the cases, we have found the same qualitative behaviour as those presented here.

## 5. Conclusions

We have analysed the conditions for the non-extinction of a resource dependent age structured population under a constant mortality harvesting strategy or proportional harvest. We have shown that the introduction of harvesting decreases the inherent net reproductive rate of a population, and for small harvesting mortality and fertility coefficients, the dynamics of the harvested population is qualitatively equivalent to the dynamics of the non harvested population with lower fertility coefficients, provided the harvesting mortality is small when compared with the resource dependent transition probabilities between age classes.

As it has been shown in section 3, a sufficient condition for a population to extinguish is that $\delta \geq \min_i \alpha_i(R^n)$, where $\delta$ is the harvesting mortality and $\alpha_i(R^n)$ is the resource dependent transition probability between age classes. This implies that proportional harvest depends on the resources available to the population.

It has been shown that the disappearance of older age classes is an indication of over exploitation. For small carrying capacity, harvesting can lead to abrupt extinction, and abrupt extinction does not necessarily occur for high harvesting mortality. On the other hand, harvesting with increasing carrying capacity can lead to the stabilisation of erratic time behaviour.

Harvesting can be a strategy in order to stabilise periodic or quasi-periodic oscillations in the number of individuals of a population. The harvesting yield of a population can be constant, periodic, quasi-periodic or chaotic. In all the cases analysed, the maximum mean temporal behaviour of the harvesting yield always occurs with constant or period-2 time behaviour of population numbers. This last conclusion derives from the numerical analysis presented in section 4.

For populations with large fertility numbers and erratic dynamics in the temporal variation of population numbers ([2]), small harvesting mortality can lead to extinction, but larger harvesting mortality can prevent extinction by avoiding over consumption of resources. The complete bifurcation diagram of survival, extinction, periodic and chaotic time behaviour of a population as a function of fertility and harvesting mortality has been obtained.

From the more technical point of view, it has been shown that if proportional harvesting drives the system through a situation where the system loses its (local) structural stability properties, the harvesting effect can strongly change the dynamics of the population. In situation where we have quasi-periodic, chaotic or even periodic behaviour this can lead to abrupt extinction, as it is summarized in figure 4.

**Acknowledgements**: This work has been partially supported by the PRAXIS XXI Project P/FIS/13161/1998 (Portugal). We would like also to acknowledge the valuable remarks of two anonymous referees of this paper.

**Figure Captions**

**Figure 1:** In grey, we show the bifurcation diagram for map (2.1) with model functions (2.4). We have plotted the asymptotic states of the total population number $N = \sum_{i=1}^{k} N_i$ as a function of the fertility coefficient $e_2$. For each parameter value, we have calculated 2500 iterates of the map and we have plotted the last 1500 iterates. The other chosen parameter values are: $k = 3$, $e_3 = 0.8$, $K = 100$, $\gamma_1 = \gamma_2 = 0.1$, $\mu = 1$ and $\beta = 1000$. For this choice of parameters, and, by (2.3), calculating $G(K)$, the non-zero period-1 fixed point exists for $e_2 > 0.34$ ($\delta = 0$). When $G(K)$ increases by increasing $e_2$, the period-1 fixed point loses stability by a flip or period doubling bifurcation and, for larger values of $e_2$, the period-2 fixed point loses stability by a Hopf bifurcation leading to quasi-periodic time behaviour. In black, we show the bifurcation diagram for map (3.1) with model functions (2.4) and harvesting mortality $\delta = 0.2$. In this case, by (3.2), the non-zero period-1 fixed point exists for $e_2 > 0.842$, and the harvested population goes extinct for $e_2 > 3.46$. Away from the bifurcation values of map (2.1), the qualitative features of map (3.1) remain unchanged. Clearly, the map (3.1) should be understood has a perturbation of map (2.1). In the limit $\delta \to 0$, both bifurcation diagrams coincide.

**Figure 2:** Bifurcation diagram for the total population number $N = \sum_{i=1}^{k} N_i$ as a function of the carrying capacity $K$ for maps (2.1), in grey, and (3.1), in black, with model functions (2.4). The parameter values are $k = 3$, $e_2 = 2.8$, $e_3 = 0.8$, $\gamma_1 = \gamma_2 = 0.1$, $\mu = 1$ and $\beta = 1000$. The decrease of the carrying capacity in the presence of harvesting can lead to abrupt extinction. The abrupt extinction occurs when the fixed points of the map

(3.1) obey the condition $R^n = \max_i \alpha_i^{-1}(\delta)$. On the other hand, harvesting with increasing carrying capacity can lead to the stabilisation of erratic time behaviour.

**Figure 3:** Bifurcation diagram of the harvesting yield $\delta N$ (black) and number of surviving individuals of the population $(1-\delta)N$ (grey) as a function of harvesting mortality $\delta$, for map (3.1) with model functions (2.4) and parameter values $k = 3$, $e_3 = 0.8$, $K = 100$, $\gamma_1 = \gamma_2 = 0.1$, $\mu = 1$ and $\beta = 1000$. The harvesting threshold of extinction is calculated by the condition $G(K, \delta) = 1$. For populations with large fertility numbers $\sum_{i=2}^{k} e_i$, small harvesting mortality can lead to extinction, but larger mortalities can stabilise the harvested population. In the case c), $e_2 = 3.46$, the population becomes extinct for harvesting mortality in the range $\delta \in [0.0168, 0.196]$ and for $\delta > 0.637$. In figures d) and e), we show the mean value in time of the harvesting yield as a function of $\delta$. In these cases, the maximum temporal mean harvesting yield occurs outside the erratic regions, when the harvesting yield is periodic.

**Figure 4:** Regions of survival and extinction for map (3.1) with $k = 3$ as a function of the harvesting mortality $\delta$ and the fertility coefficient $e_2$. The fixed parameter values are: $e_3 = 0.8$, $K = 100$, $\gamma_1 = \gamma_2 = 0.1$, $\mu = 1$ and $\beta = 1000$. The upper curve is calculated from (3.3) and is asymptotic to $\delta = \alpha_1(K) = 10/11$. The lower curve is calculated numerically. The bifurcation diagrams of figures 1 and 2 have been calculated along the dotted lines. In the region of survival, we show the regions where the population has a periodic stable steady state and erratic time behaviour. The extinction regions correspond to the case where the orbits of map (3.1) obey to the condition $R^n \leq \max_i \alpha_i^{-1}(\delta)$, for some $n$. In the region we label chaotic, it is difficult to distinguish by simple qualitative analysis if the behaviour of orbits of the dynamics is quasi-periodic or chaotic. However, in any of these two possible dynamical situations, the temporal evolution is erratic.

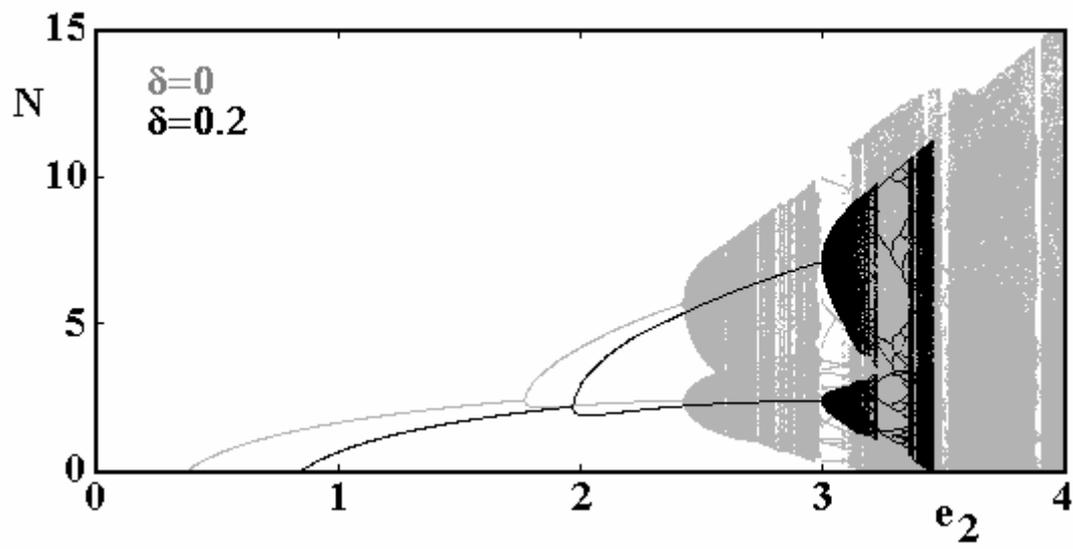

**FIGURE 1**

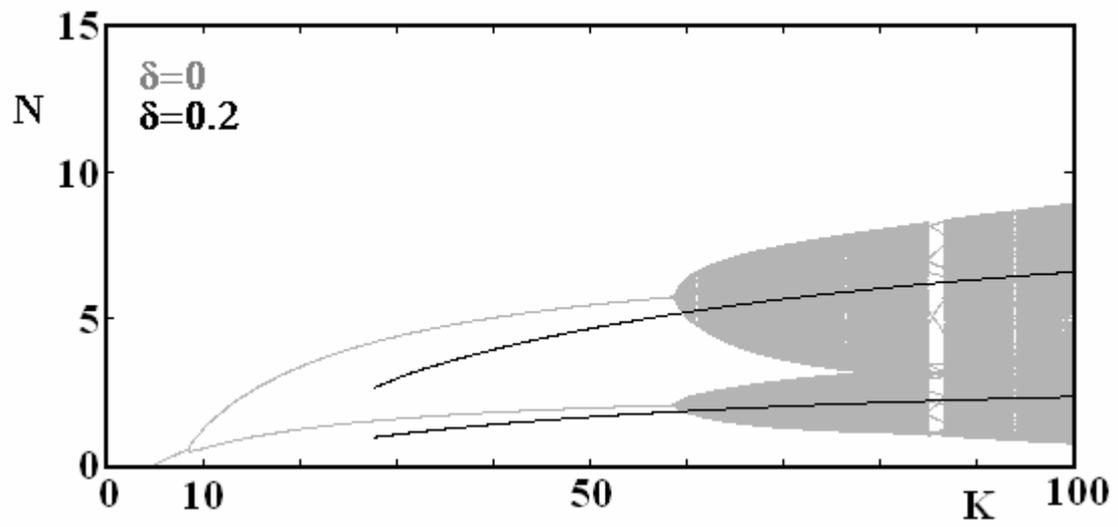

**FIGURE 2**

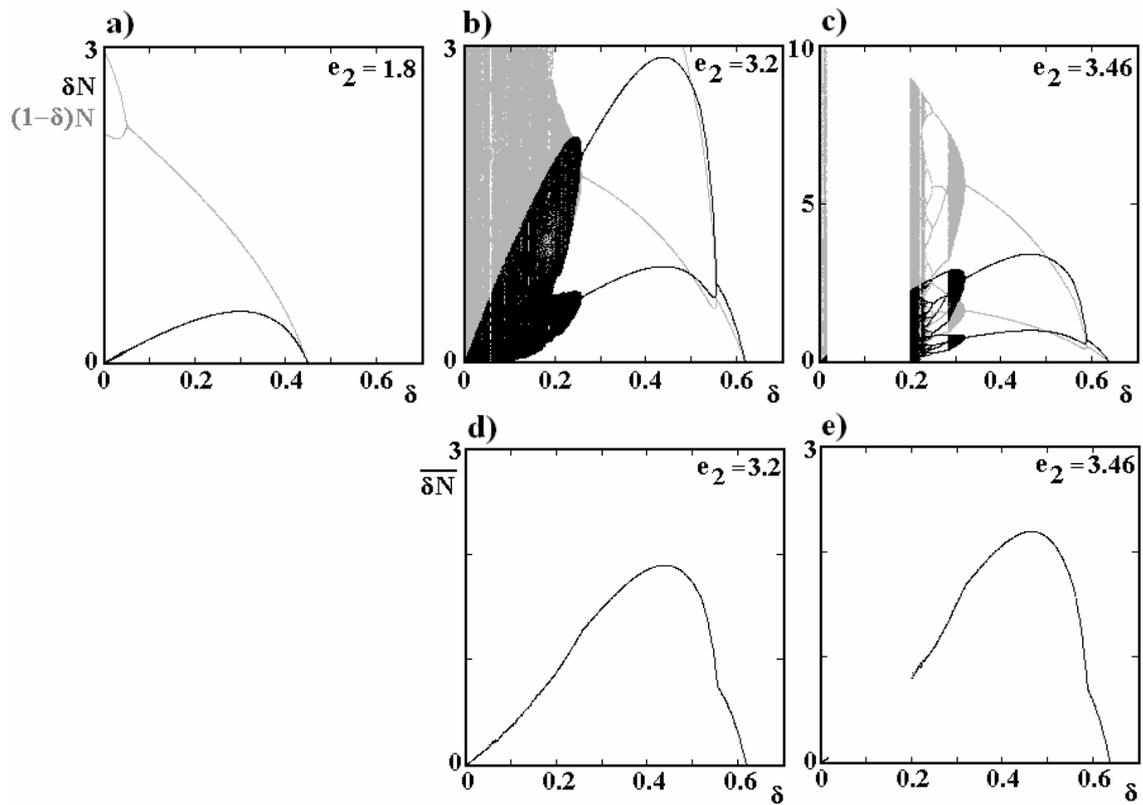

**FIGURE 3**

Dilão, Domingos and Shahverdiev

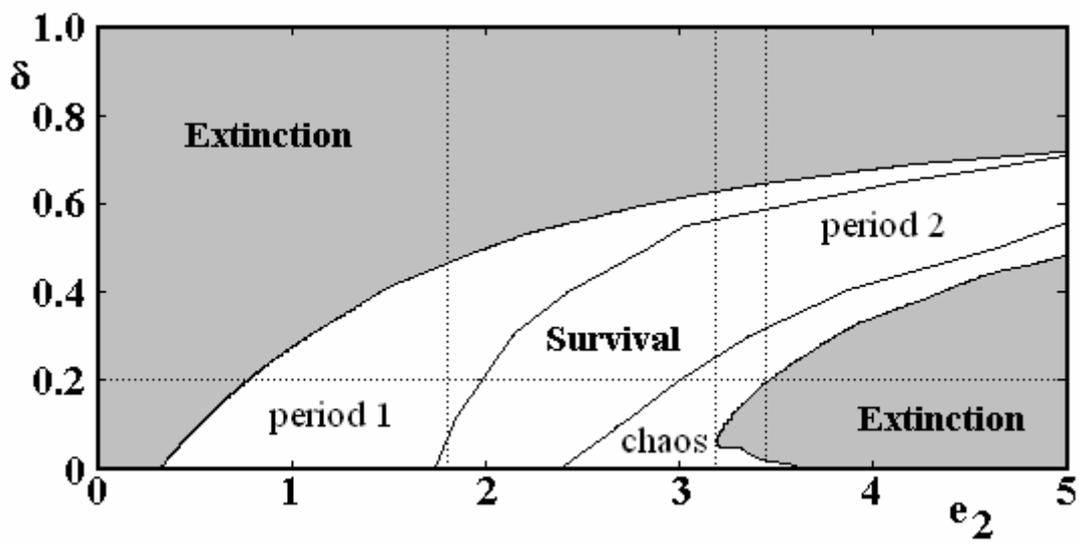

**FIGURE 4**

Dilão, Domingos and Shahverdiev